\documentclass[12pt,a4paper]{article}

\usepackage[utf8]{inputenc}
\usepackage[T1]{fontenc}
\usepackage[english]{babel}
\usepackage{amsmath, amssymb}
\usepackage{graphicx}
\usepackage[hidelinks]{hyperref}
\usepackage{booktabs}
\usepackage{listings}
\usepackage{xcolor}
\usepackage{geometry}
\usepackage[expansion=false]{microtype}
\usepackage{float}
\usepackage{url}
\usepackage[numbers,square]{natbib}
\usepackage{multicol}
\usepackage{tikz}
\usepackage{pgfplots}
\pgfplotsset{compat=1.16}
\usetikzlibrary{shapes.geometric, arrows.meta, positioning, fit, backgrounds, calc}

\definecolor{codegray}{rgb}{0.95,0.95,0.95}
\definecolor{codegreen}{rgb}{0,0.45,0}
\definecolor{codeblue}{rgb}{0,0,0.55}
\definecolor{codepurple}{rgb}{0.45,0,0.45}

\lstdefinestyle{cstyle}{
  backgroundcolor=\color{codegray},
  commentstyle=\color{codegreen}\itshape,
  keywordstyle=\color{codeblue}\bfseries,
  stringstyle=\color{codepurple},
  basicstyle=\ttfamily\footnotesize,
  breaklines=true, keepspaces=true,
  numbers=left, numberstyle=\tiny\color{gray},
  showstringspaces=false, tabsize=4,
  language=C, frame=single, rulecolor=\color{lightgray}
}

\title{\textbf{Renting the Cracking Machine with a Cost-and-Time Analysis of Exhaustive DES-56 Key Search in the Cloud}}

\author{
  Gonzalo Sharif Curi Martínez\\
  \small Instituto Tecnológico de Buenos Aires (ITBA)\\
  \small \texttt{gcurimartinez@itba.edu.ar}
  \and
  Rodrigo Ramele\\
  \small Instituto Tecnológico de Buenos Aires (ITBA)\\
  \small \texttt{rramele@itba.edu.ar}
}
\date{July 26, 2026}

\begin{document}
\maketitle

\begin{abstract}
The Data Encryption Standard (DES), with its 56-bit key, has been considered cryptographically broken since 1998.
However, a concrete, reproducible measurement of the cost and time required to perform an exhaustive key search using
today's commodity cloud infrastructure has not been widely reported in recent literature.
In this paper we present a distributed brute-force system built on AWS EC2 that partitions the $2^{56}$ keyspace
across 37 \texttt{c6i.2xlarge} instances running a C/OpenMP worker, achieving a measured throughput of
$2.91$\,M\,keys/s per instance (${\sim}108 \times 10^6$ keys/s aggregate).
We conduct 15 independent trials covering keyspace offsets from $10^6$ to $1.5 \times 10^{10}$ keys, measuring
wall-clock time and monetary cost per trial. For small offsets ($\leq 10^8$), total time is dominated by AWS
instance boot latency (${\approx}\;90$\,s), yielding a mean of $116.8\pm20.1$\,s at \$0.41 per attack. For larger
offsets the search time dominates and grows linearly: a key at offset $1.5{\times}10^{10}$ requires
~${\approx} \; 87$ minutes and \$18. At the measured aggregate throughput of 108\,M\,keys/s, exhausting the full
$2^{56}$ keyspace with these 37 instances would take ${\approx} \; 21$ years; however, because the workload is
embarrassingly parallel and cloud capacity is elastic, the same search can be traded for money almost linearly.
Extrapolating our measured cost, a complete exhaustive search would cost ${\approx} \; \$1.2$M and, with a
sufficiently large fleet, could be completed in about one day. The system is thus practical for
bounded-subspace attacks at negligible cost, and full DES exhaustion, while expensive, is firmly within reach
of a well-funded attacker using only commodity cloud resources.
\end{abstract}

\textbf{Keywords:} DES, brute-force, cloud computing, AWS EC2, OpenMP

\section{Introduction}

The Data Encryption Standard (DES) \citep{des1977} was the dominant symmetric cipher from 1977 to the late 1990s.
Its principal weakness is its 56-bit effective key length ($2^{56} \approx 7.2\times10^{16}$ possible keys), which
has been considered cryptographically insufficient since the late 1970s \citep{diffie1977}. Prior work has
demonstrated exhaustive key search attacks using dedicated ASICs and FPGAs (see Section~\ref{sec:priorwork}), but
these approaches required significant upfront investment in specialized, non-reusable hardware.

A natural follow-up question is: how expensive and how fast is the same attack with today's commodity cloud
resources, available to anyone with a credit card? Cloud computing has dramatically changed the economics of
computation: large clusters of CPUs can be rented for seconds at a time, charged by the second, and terminated
when done. Distributed, embarrassingly parallel workloads -- such as exhaustive key search -- are ideally suited
for this model \citep{cloudcracking2010}.

In this work we implement a full DES brute-force system on AWS EC2 and conduct 15 independent experiments
covering keyspace offsets from $10^6$ to $1.5\times10^{10}$ keys. We report throughput, time-to-solution, and
cost across both a boot-dominated and a search-dominated regime, providing an empirical, reproducible answer to
the title question.

\section{Background}

\subsection{The DES Algorithm}

DES is a 64-bit block cipher based on a 16-round Feistel network \citep{des1977}. Each round applies the function
$f(R, K_i)$, which consists of: (1) expansion $E$ ($32 \to 48$ bits), (2) XOR with subkey $K_i$, (3) substitution
through 8 S-boxes ($48 \to 32$ bits), and (4) fixed permutation $P$. The 56-bit key (the remaining 8 bits are
parity) generates 16 subkeys via the key schedule (PC-1 and PC-2 permutations and left-circular shifts).

\subsection{Prior Work on DES Key Search}
\label{sec:priorwork}

The cost of breaking DES has been revisited as technology evolved. Diffie and Hellman \citep{diffie1977}
estimated in 1977 that a \$20M machine could search the keyspace in about a day; Wiener \citep{wiener1993} later
proposed a \$1M design recovering a key in 3.5 hours on average. The definitive empirical result came in 1998,
when the EFF built Deep Crack for \$250,000---1,856 custom ASICs that exhausted the keyspace in under 56 hours
\citep{eff1998}. Commodity FPGAs then lowered the barrier: COPACOBANA (2006, 120 FPGAs, under \$10,000) averaged
8.7 days \citep{copacobana2006}, and its successor RIVYERA pushed this below a day \citep{sciengines2008}. More
recently, Cui and Zhang \citep{cui2025} introduced algorithmic optimizations (composite S/P-box tables, partial
matching) reducing per-key cost by ${\approx}16\times$, but without addressing deployment cost on real
infrastructure.

Notably, none of these works measures the cost and time of a DES key search on commodity cloud infrastructure,
where there is no upfront capital cost and the attacker pays only for the seconds used. This paper provides the
first empirical answer to that question.

\section{Materials and Methods}

\subsection{System Architecture}

The system comprises three components:

\textbf{Coordinator} (Python + boto3, local machine): partitions the $2^{56}$ keyspace into $W$ disjoint
ranges and launches $W$ EC2 instances, each receiving its assigned range via a \textit{user-data}
script.\footnote{The worker enumerates the 56 effective key bits and inserts the 8 parity bits before the DES
key schedule, so the counter ranges over the $2^{56}$ distinct effective keys with no redundancy. Each trial
therefore tests a unique effective key.} With $W = 37$, worker $i$ covers $[i\delta, (i+1)\delta)$ where
$\delta = \lfloor 2^{56}/37 \rfloor \approx 1.95\times10^{15}$ keys (the last worker extends to $2^{56}$). It
collects results over the Simple Queue Service (SQS), a fully managed message-queue service in which producers
push messages that a consumer later retrieves, using long-polling (a request that blocks for up to a 20\,s
timeout until a message arrives, avoiding busy-waiting); and it uploads a stop flag to the Simple Storage
Service (S3), a managed object store, once a matching key (KPA) or a candidate above the score threshold (COA)
is found.

\textbf{Workers} (C/OpenMP, AWS C6i-family instances \texttt{c6i.2xlarge}, 8 vCPUs each): each worker receives
$[k_{\text{start}}, k_{\text{end}})$ and iterates over every key in that range. OpenMP distributes the sub-range
across the 8 available vCPUs, with one independent top-$K$ candidate list per thread to avoid lock contention.
Workers poll S3 every 30\,s for the stop flag.

\textbf{AWS managed services}: the design uses SQS for worker-to-coordinator communication (workers push
results and heartbeats into the queue) and S3 for the stop flag (workers poll the object). This split reflects
the two access patterns: SQS is well suited to many producers sending to one consumer, whereas a single SQS
message can only be consumed once, so it cannot broadcast a stop signal to all 37 workers; an S3 object, which
every worker can read independently, serves that broadcast role instead.

\begin{figure}[H]
\centering
\begin{tikzpicture}[
  box/.style={rectangle, rounded corners=5pt, draw, thick,
              minimum width=3.2cm, minimum height=1.1cm,
              align=center, font=\small},
  coord/.style={box, fill=gray!15},
  svc/.style={box, fill=teal!15},
  worker/.style={box, fill=violet!12},
  arr/.style={-{Stealth[length=5pt]}, thick},
  darr/.style={-{Stealth[length=5pt]}, thick, dashed},
  barr/.style={-{Stealth[length=5pt]}, very thick},
  lbl/.style={font=\scriptsize, fill=white, inner sep=2pt},
]

\node[coord]  (coord) at (0,   0) {Coordinator\\\scriptsize Python + boto3};
\node[worker] (ec2)   at (6.5, 0) {37 × EC2 \texttt{c6i.2xlarge}\\\scriptsize C/OpenMP};
\node[svc]    (sqs)   at (3.2, -2.8) {SQS Queue\\\scriptsize results + heartbeats};
\node[svc]    (s3)    at (3.2,  2.8) {S3 \texttt{stop\_flag}};

\begin{scope}[on background layer]
  \node[draw, dashed, rounded corners=8pt, thick, gray,
        fit=(ec2)(sqs)(s3),
        inner sep=16pt,
        label={[font=\small, gray]above:AWS}] {};
\end{scope}

\draw[arr]
  (coord.east) -- (ec2.west)
  node[lbl, pos=0.5, above]{\textbf{1} run\_instances};

\draw[darr]
  (ec2.south) -- ++(0,-0.5) -| (sqs.east)
  node[lbl, pos=0.25, right]{\textbf{2} heartbeat};

\draw[barr]
  (ec2.south west) -- ++(0,-0.3) -| (sqs.north east)
  node[lbl, pos=0.6, left]{\textbf{3} candidates};

\draw[arr]
  (sqs.west) -- ++(-0.5,0) |- (coord.south)
  node[lbl, pos=0.3, below]{\textbf{4} long poll (20\,s)};

\draw[barr]
  (coord.north) -- ++(0,0.5) -| (s3.west)
  node[lbl, pos=0.6, above]{\textbf{5} upload stop\_flag};

\draw[darr]
  (s3.east) -- ++(0.5,0) |- (ec2.north)
  node[lbl, pos=0.3, above]{\textbf{6} poll every 30\,s};

\end{tikzpicture}
\caption{System architecture. Numbers indicate the sequence of events during an attack.
Solid arrows: active communication. Dashed arrows: polling.
Double arrows (\textbf{3}, \textbf{5}): primary data path.}
\label{fig:arch}
\end{figure}
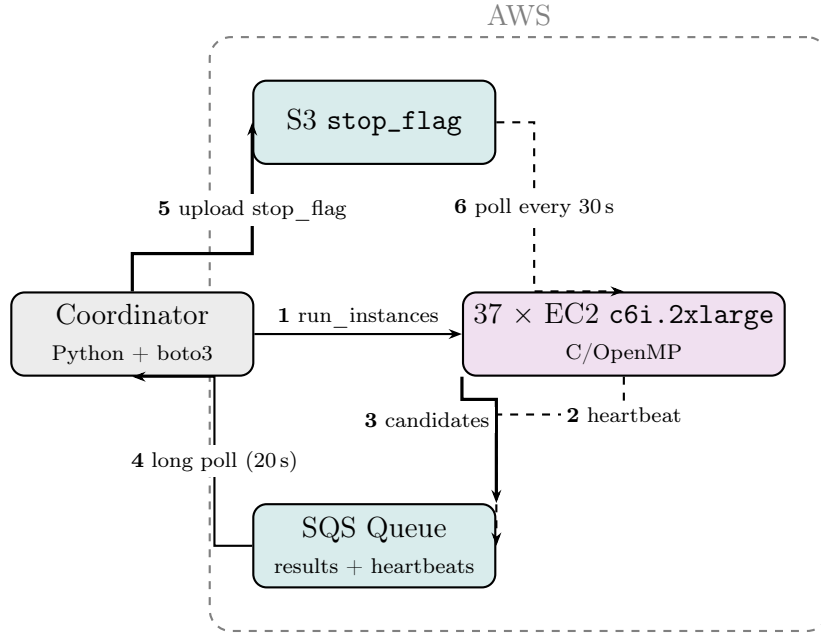

Figure~\ref{fig:arch} shows the full communication topology.

\subsection{Ciphertext-Only Scoring}

The system implements two attack modes. In known-plaintext attack (KPA) mode, the attacker possesses a
matching plaintext--ciphertext pair, and each candidate key is verified by an exact comparison --- a key is
correct if and only if it reproduces the known plaintext. In ciphertext-only attack (COA) mode, only the
ciphertext is available, and the correct key must be identified statistically by recognizing when the decrypted
output resembles meaningful text. This paper focuses on COA, which is both the more challenging and the more
realistic scenario: it requires no prior knowledge of the plaintext and is the mode an attacker would face when
intercepting encrypted traffic.

In COA mode, the correct key is identified by a log-frequency score over a Spanish letter-frequency model. Each
byte $b$ receives a log-weight $w(b)$ derived from Spanish unigram frequencies (printable characters range from
about $-2.0$ for the most common symbols to $-6.2$ for the rarest); any non-printable byte receives a $-100$
penalty. The score of a candidate key is
\begin{equation}
  \text{score}(k) = \sum_{b \in \text{dec}(k)} w(b)
  \label{eq:score}
\end{equation}
Since every candidate first passes the printable-ASCII filter, the $-100$ floor acts as the rejection mechanism
rather than a term in the sum. Legitimate Spanish text scores roughly $-2.5$ to $-3.5$ per byte, while keys that
pass the ASCII filter by chance average ${\approx} -4.7$. The early-exit threshold of $-4.0 \times n_{\text{bytes}}$
(evaluated on the full multi-block score) separates the two cleanly.

\subsection{DES Implementation and Optimizations}

A naive bit-by-bit DES implementation achieves ${\sim}125{,}000$ keys/s. We apply four independent
optimizations, building on classic software-DES techniques for combining the S-box and permutation stages
\citep{biham1997,kwan2000}:

\textbf{SP-tables.} The permutation $P$ is linear: $P(A \oplus B) = P(A) \oplus P(B)$. Since the outputs of the
8 S-boxes occupy disjoint bit positions in the 32-bit word, $P(\text{sbox\_out}) = \bigoplus_{s=0}^{7} P(X_s)$.
Each $P(X_s)$ depends only on the 6-bit input to S-box $s$ ($2^6 = 64$ possibilities). We precompute
$SP[s][i] = P(\text{``}S_s(i)\text{ in its position''})$, a $8 \times 64 \times 4 = 2048$ byte table that fits
entirely in L1 cache. The full Feistel round reduces to 8 table lookups and 7 bitwise ORs.

\textbf{Byte-indexed LUTs for IP, IP$^{-1}$, PC-1, PC-2.} Every bit permutation is linearly decomposable over
bytes. For example, $IP(x) = \bigvee_{k=0}^{7} \text{ip\_lut}[k][(x \gg (56-8k)) \,\&\, \text{0xFF}]$. The four
tables total ${\sim}62$~KB ($8 \times 256 \times 8$ bytes for IP, IP$^{-1}$, and PC-1; $7 \times 256 \times 8$
bytes for PC-2, whose input is 56 bits), fitting in L2 cache.

\textbf{Direct bit-shift expansion $E$.} The 48-bit expansion of $R$ is computed with 8 constant bit-shift
expressions derived from the DES standard, eliminating the 48-iteration loop entirely. In fact, the expansion
is fused with the subkey XOR and SP-table indexing: no explicit 48-bit $E(R)$ value is ever materialized.

\textbf{Early rejection (COA mode).} For a random key, the probability that all 8 bytes of one decrypted block
fall in the printable ASCII range [0x20, 0x7E] (95 of 256 values) is $(95/256)^8 \approx 0.036\%$. Blocks are
tested one by one; decryption halts on the first non-printable byte. In practice, 99.96\% of keys are rejected
after a single block, making the effective cost per key ${\approx} 1\times$ one block rather than $10\times$
(the default block count for scoring).

Collectively, these optimizations reduce the Feistel round to 8 table lookups and 7 bitwise ORs for the S-box
and permutation $P$, eliminate all bit-level loops for IP, IP$^{-1}$, PC-1, and PC-2, and remove the
48-iteration expansion loop entirely. Combined with \texttt{gcc -O3} and OpenMP across 8 vCPUs, the final
measured throughput is 2.91\,M\,keys/s per instance in COA mode with 10 ciphertext blocks (derived by linear
regression over trials 11--15; see Section~\ref{sec:results}).

\subsection{Experimental Design}

To measure the system's performance empirically and address the title question, we conduct 15 independent
trials organized in two groups:

\begin{enumerate}
\item \textbf{Group A (trials 1--10, boot-dominated):} We fix a known plaintext (88-byte Spanish poem encoded
as 11 DES blocks, of which the first 10 are used for scoring) and construct 10 keys as
$k_i = k_{\text{start}}^{(w_i)} + \delta_i$, with the worker indices $w_i$ and offsets $\delta_i$ chosen to
increase monotonically across trials (workers $0, 4, 8, \ldots, 36$; offsets from $10^6$ to $10^8$). This
spreads the trials across the keyspace while keeping the search component small, characterizing the system's
baseline performance when infrastructure latency dominates.

\item \textbf{Group B (trials 11--15, search-dominated):} Five additional trials with increasing offsets
$\delta_i \in \{10^9, 3\times10^9, 5\times10^9, 10^{10}, 1.5\times10^{10}\}$, placed in workers 2, 10, 18, 26,
and 34 respectively. This group allows empirical measurement of per-worker throughput and characterizes how
total time scales with keyspace position.

\item For each trial we record: (a) total wall-clock time from coordinator launch to full instance
termination, and (b) AWS cost.

\item We report mean $\mu$ and standard deviation $\sigma$ for Group A, and per-instance throughput derived
from Group B.
\end{enumerate}

All trials use 37 on-demand \texttt{c6i.2xlarge} instances (\$0.34/h) in \texttt{us-east-1}, the same
pre-compiled AMI, and 10 ciphertext blocks. The system is validated beforehand against the FIPS-46 test vector
(key: \texttt{133457799BBCDFF1}, plaintext: \texttt{0123456789ABCDEF}, expected ciphertext:
\texttt{85E813540F0AB405}).

\section{Results}
\label{sec:results}

\subsection{Per-Trial Measurements}

Table~\ref{tab:trials} shows all 15 trial results, split into two groups: trials 1--10 with small offsets
($\leq 10^8$) where boot latency dominates, and trials 11--15 with larger offsets ($10^9$--$1.5\times10^{10}$)
where search time dominates. All times are wall-clock times from coordinator launch to full instance
termination. Cost is computed as $37 \times \$0.34/\text{h} \times t_{\text{total}}$ and represents an upper
bound: since instances are launched sequentially over ${\approx}55$\,s, the average instance runs for
$t_{\text{total}} - 27.5$\,s, making the true cost slightly lower (by ${\approx}\$0.03$--\$0.06 per trial).

\begin{table}[H]
\centering
\caption{Results of 15 independent trials with increasing offsets. All times include instance boot
(${\approx}90$\,s), key search, and shutdown. Keys are enumerated over the effective 56-bit keyspace; the
recovered key matches the original up to parity bits (see Section~\ref{sec:parity}).}
\label{tab:trials}
\begin{tabular}{@{}clllc@{}}
\toprule
Trial & Recovered key (hex) & Offset & $t_{\text{total}}$ & Cost (\$) \\
\midrule
1  & \texttt{00000000007a0880} & $1\times10^6$ & 101\,s & 0.35 \\
2  & \texttt{1ad63e2250701818} & $5\times10^6$ & 105\,s & 0.37 \\
3  & \texttt{36ac7c44a0e03030} & $1\times10^7$ & 100\,s & 0.35 \\
4  & \texttt{5282ba66f6b274c8} & $2\times10^7$ & 106\,s & 0.37 \\
5  & \texttt{6e58f88a4c84ba60} & $3\times10^7$ & 101\,s & 0.35 \\
6  & \texttt{8a3036acac1c58f8} & $5\times10^7$ & 106\,s & 0.37 \\
7  & \texttt{a60674d00ab2f890} & $7\times10^7$ & 131\,s & 0.46 \\
8  & \texttt{c0dcb2f260863e28} & $8\times10^7$ & 127\,s & 0.44 \\
9  & \texttt{dcb2f214b65882c0} & $9\times10^7$ & 129\,s & 0.45 \\
10 & \texttt{f88a30380c2ac858} & $1\times10^8$ & 162\,s & 0.57 \\
\multicolumn{2}{l}{$\mu \pm \sigma$ (trials 1--10)} & $\leq 10^8$ & $116.8 \pm 20.1$\,s & $0.408 \pm 0.070$ \\
\midrule
11 & \texttt{0cea9e985edc9ccc} & $1\times10^9$   & 6m 35s     & 1.38 \\
12 & \texttt{44981aec6aa4c2fc} & $3\times10^9$   & 17m 06s    & 3.59 \\
13 & \texttt{7c449840766cea2c} & $5\times10^9$   & 27m 35s    & 5.78 \\
14 & \texttt{b2f214aaaeb8885c} & $1\times10^{10}$ & 56m 39s   & 11.88 \\
15 & \texttt{ea9e9214e804268c} & $1.5\times10^{10}$ & 1h 26m 31s & 18.14 \\
\bottomrule
\end{tabular}
\end{table}

\subsection{Analysis}

The results reveal two distinct regimes determined by the relationship between offset and throughput.

\textbf{Boot-dominated regime (trials 1--10, offset $\leq 10^8$).} For these small offsets the actual key
search takes at most a few tens of seconds (e.g. ${\approx}34$\,s for the largest, offset $10^8$), so total
time ($116.8 \pm 20.1$\,s) is still largely explained by AWS provisioning (${\approx}90$\,s) and instance
shutdown (${\approx}15$\,s average, up to 30\,s worst case due to the S3 poll interval). The variance across
trials in this group reflects both AWS boot/shutdown jitter and the growing search component as the offset
increases.

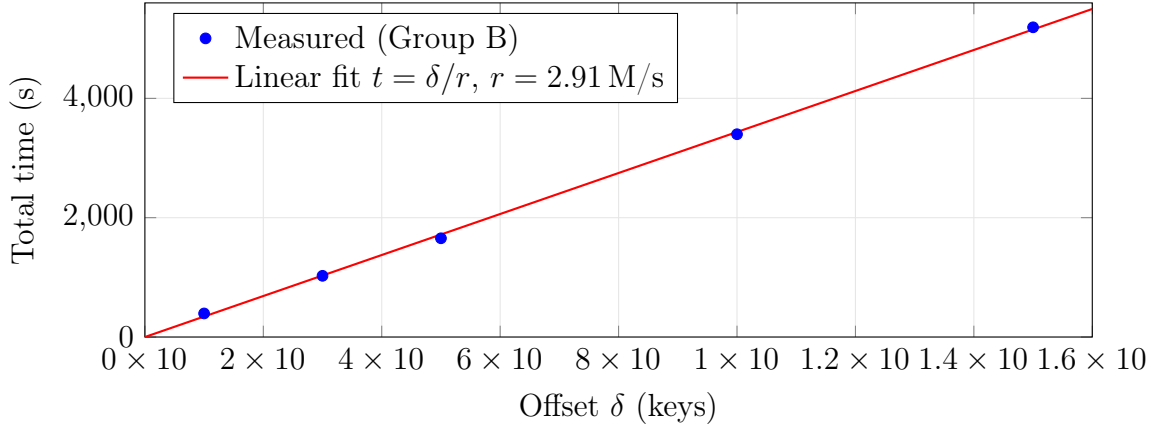
\begin{figure}[H]
\centering
\begin{tikzpicture}
\begin{axis}[
  width=0.85\textwidth, height=6cm,
  xlabel={Offset $\delta$ (keys)},
  ylabel={Total time (s)},
  xmin=0, xmax=1.6e10, ymin=0, ymax=5600,
  grid=both, grid style={gray!20},
  legend pos=north west,
  legend cell align=left,
  scaled x ticks=false,
  x tick label style={/pgf/number format/sci, /pgf/number format/sci generic={mantissa sep=\times,exponent={10^{#1}}}},
]
\addplot[only marks, mark=*, mark size=2pt, blue] coordinates {
  (1e9, 395) (3e9, 1026) (5e9, 1655) (10e9, 3399) (15e9, 5191)
};
\addlegendentry{Measured (Group B)}
\addplot[red, thick, domain=0:1.6e10, samples=2] {x/2.91e6};
\addlegendentry{Linear fit $t=\delta/r$, $r=2.91$\,M/s}
\end{axis}
\end{tikzpicture}
\caption{Search-dominated regime: total time grows linearly with offset. The slope gives a per-worker
throughput of $r = 2.91$\,M\,keys/s ($R^2 = 0.9994$).}
\label{fig:linear}
\end{figure}

\textbf{Search-dominated regime (trials 11--15, offset $\geq 10^9$).} For larger offsets the search time
dominates and grows linearly with offset, as expected. We fit a linear model $t_{\text{total}} = \delta/r + c$
to the five Group B trials, where $\delta$ is the offset and $r$ is the per-worker throughput. The fit yields
$r = 2.91$\,M\,keys/s per worker (${\approx}108$\,M\,keys/s aggregate across 37 instances), with $R^2 = 0.9994$.
Figure~\ref{fig:linear} shows the fit against the measured points. The near-perfect linearity confirms that the
system behaves as a constant-throughput search engine, and that the throughput is stable across an order of
magnitude in offset.

\textbf{Important caveat on scope.} These results should not be interpreted as evidence that DES can be broken
in minutes for an arbitrary key. The trials cover only a small fraction of the $2^{56}$ keyspace: the largest
offset tested ($1.5{\times}10^{10}$) represents just $2{\times}10^{-5}$\% of the full keyspace. At the measured
aggregate throughput of 108\,M\,keys/s, a complete exhaustive search of the $2^{56}$ effective keyspace would
require $2^{56} / 108{\times}10^6 \approx 6.7{\times}10^8$\,s, or roughly 21 years in the worst case (the
expected time is half that, ${\approx}10.6$ years).

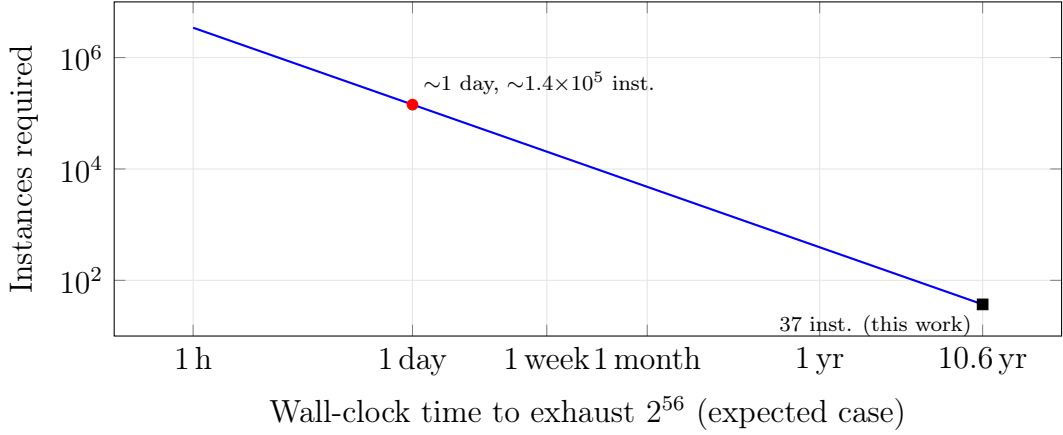
\begin{figure}[H]
\centering
\begin{tikzpicture}
\begin{axis}[
  width=0.85\textwidth, height=6cm,
  xlabel={Wall-clock time to exhaust $2^{56}$ (expected case)},
  ylabel={Instances required},
  xmode=log, ymode=log,
  ymin=10, ymax=1e7,
  xtick={0.04166667, 1, 7, 30, 365, 3869},
  xticklabels={1\,h, 1\,day, 1\,week, 1\,month, 1\,yr, 10.6\,yr},
  grid=both, grid style={gray!20},
]
\addplot[blue, thick, domain=0.04166667:3869, samples=100] { (2^56/2)/(2.91e6 * x*86400) };
\addplot[only marks, mark=*, red, mark size=2pt] coordinates {(1, 1.43e5)};
\node[anchor=south west, font=\scriptsize] at (axis cs:1,1.43e5) {${\sim}1$ day, ${\sim}1.4{\times}10^5$ inst.};
\addplot[only marks, mark=square*, mark size=2pt, black] coordinates {(3869, 37)};
\node[anchor=north east, font=\scriptsize] at (axis cs:3869,37) {37 inst. (this work)};
\end{axis}
\end{tikzpicture}
\caption{Cost/time scaling for full $2^{56}$ exhaustion (expected case). The horizontal axis is the target
wall-clock time (log scale), reaching 10.6 years with our 37-instance fleet at the far right; the vertical axis
is the number of instances needed to hit that time. Because total worker-hours are fixed, the total monetary
cost is approximately constant (${\approx}\$1.2$M) along the whole curve --- only the wall-clock time changes.
Reaching a one-day search requires ${\sim}1.4\times10^5$ instances.}
\label{fig:scaling}
\end{figure}

However, because exhaustive search is embarrassingly parallel and cloud capacity is elastic, this time can be
traded for money almost linearly: doubling the number of instances halves the wall-clock time at
(approximately) constant total cost. Figure~\ref{fig:scaling} shows this trade-off. The total cost of a full
worst-case exhaustion is essentially fixed regardless of fleet size:
\begin{equation}
  \text{Cost}_{\text{full}} \approx \frac{2^{56}}{r_{\text{worker}} \cdot 3600} \times \$0.34/\text{h}
  \approx \$2.3\text{M} \quad (\text{worst case}),
  \label{eq:costfull}
\end{equation}
or ${\approx}\$1.2$M for the expected case. Within this fixed budget, the attacker chooses the wall-clock time
by scaling the fleet: ${\approx}1.4\times10^5$ instances complete the expected-case search in about a day, and
${\approx}10^6$ instances in a few hours. Thus, while a single 37-instance run is only practical for bounded
subspaces, full DES exhaustion is a question of budget rather than feasibility for a well-funded attacker.

\subsection{Note on DES Parity Bits}
\label{sec:parity}

Each byte of the 64-bit DES key reserves its least significant bit for parity, leaving 56 effective bits. Our
worker enumerates these 56 bits with all parity bits zeroed, producing the \emph{canonical} form of each key;
any two keys differing only in parity bits encrypt identically. For example, Trial~1 targets effective-key
index $10^6$, whose canonical form \texttt{00000000007a0880} is exactly what the search recovers. All 15 trials
recover the canonical key matching their target index, confirmed by successful plaintext recovery in every case.

\section{Conclusion}

We presented a distributed brute-force system for DES on 37 AWS EC2 \texttt{c6i.2xlarge} instances, measuring an
aggregate throughput of 108\,M\,keys/s ($2.91$\,M per instance, $R^2 = 0.9994$). Across 15 trials we observe two
regimes: boot-dominated (offset $\leq 10^8$, ${\approx}2$\,min, ${\approx}\$0.41$) and search-dominated, where
time grows linearly with offset (offset $1.5\times10^{10}$, ${\approx}87$\,min, ${\approx}\$18$).

For bounded subspaces the attack is trivially cheap: any key within a known, manageable fraction of the
keyspace is recovered in minutes for well under a dollar. Full exhaustion is more expensive but not infeasible:
at our measured throughput it costs ${\approx}\$1.2$M in the expected case, and thanks to the elastic,
embarrassingly parallel nature of the workload, the attacker freely trades money for time ---
${\sim}1.4\times10^5$ instances would complete the search in about a day. DES thus offers no meaningful
security today: bounded-key scenarios fall for pocket change, and full recovery is a budget question, not a
technical barrier. Compared to COPACOBANA (2006, \$10,000, 8.7-day average) \citep{copacobana2006}, commodity
cloud infrastructure offers a dramatically lower barrier to entry with no upfront hardware cost.

\textbf{Future work: GPU acceleration.} Our CPU-based workers are far from the most cost-effective option. GPU
implementations of DES exhaustive search report throughputs orders of magnitude higher per device: Ahmadzadeh
et al. \citep{ahmadzadeh2018} demonstrate a high-performance, energy-efficient key search on DES-like
cryptosystems, and a more recent study reports ${\approx}3.87$ billion keys/s for DES on a single RTX 3070 GPU,
outperforming FPGA clusters such as COPACOBANA in price-performance \citep{adomnicai2022}. A single modern GPU
thus matches over a thousand of our CPU instances; porting the worker to CUDA would lower both the wall-clock
time and the dollar cost of full exhaustion by one to two orders of magnitude, and is the natural next step for
this work.

\bibliographystyle{unsrtnat}

\end{document}